\documentclass[fleqn,usenatbib]{mnras}

\usepackage{newtxtext,newtxmath}
\usepackage[T1]{fontenc}

\DeclareRobustCommand{\VAN}[3]{#2}
\let\VANthebibliography\thebibliography
\def\thebibliography{\DeclareRobustCommand{\VAN}[3]{##3}\VANthebibliography}

\usepackage{graphicx}	% Including figure files
\usepackage{amsmath}	% Advanced maths commands
\usepackage{xcolor}
\usepackage{CJK}
\newcommand{\mdotz}{$\dot{M}_{\scriptscriptstyle Z}$}
\newcommand{\caii}{\mbox{Ca II}}
\newcommand{\taucool}{$\tau_{\rm cool}$}
\newcommand{\tauca}{$\tau_{\rm Ca}$}

\title[Planetesimal accretion in old WDs]{No evidence for a strong decrease of planetesimal accretion in old white dwarfs}

\author[S. Blouin \& S. Xu]{
Simon Blouin$^{1}$\thanks{E-mail: sblouin@uvic.ca}
and Siyi Xu$^{2}$ \begin{CJK*}{UTF8}{gbsn}(许\CJKfamily{bsmi}偲\CJKfamily{gbsn}艺\end{CJK*})
\\
$^{1}$Department of Physics and Astronomy, University of Victoria, Victoria, BC V8W 2Y2, Canada\\
$^{2}$Gemini Observatory/NSF's NOIRLab, 670 N. A'ohoku Place, Hilo, Hawaii, 96720, USA
}

\date{Accepted XXX. Received YYY; in original form ZZZ}

\pubyear{2021}

\begin{document}
\label{firstpage}
\pagerange{\pageref{firstpage}--\pageref{lastpage}}
\maketitle
\begin{abstract}
A large fraction of white dwarfs are accreting or have recently accreted rocky material from their planetary systems, thereby ``polluting'' their atmospheres with elements heavier than helium. In recent years, the quest for mechanisms that can deliver planetesimals to the immediate vicinity of their central white dwarfs has stimulated a flurry of modelling efforts. The observed time evolution of the accretion rates of white dwarfs through their multi-Gyr lifetime is a crucial test for dynamical models of evolved planetary systems. Recent studies of cool white dwarfs samples have identified a significant decrease of the mass accretion rates of cool, old white dwarfs over Gyr timescales. Here, we revisit those results using updated white dwarf models and larger samples of old polluted H- and He-atmosphere white dwarfs. We find no compelling evidence for a strong decrease of their time-averaged mass accretion rates for cooling times between 1 and 8 Gyrs. Over this period, the mass accretion rates decrease by no more than a factor of the order of 10, which is one order of magnitude smaller than the decay rate found in recent works. Our results require mechanisms that can efficiently and consistently deliver planetesimals inside the Roche radius of white dwarfs over at least 8 Gyrs.
\end{abstract}


\section{Introduction}
\label{sec:intro}
After evolving through the main sequence, the outer envelopes of stars expand on the red and asymptotic giant branches, resulting in the engulfment of their inner planetary systems \citep{villaver2007,schroder2008}. But the fate of the outer planetary system is different. While some objects might be ejected, a large fraction of the outer planetary system can survive the post-main sequence evolution and remain in bound orbits for several Gyrs after the star has entered the white dwarf cooling track \citep{duncan1998,debes2002,villaver2007,veras2016b,vanderburg2020,blackman2021}. 

Observational evidence of planetary systems around white dwarfs abounds. Almost two thousand white dwarfs \citep{dufour2017} are known to bear the spectral signature of rock-forming elements (e.g., O, Mg, Si, Ca, Fe) in their otherwise pristine H or He atmospheres. Given the efficient gravitational settling at play in white dwarfs, those heavy elements should sink out of view in a matter of days to Myrs \citep{paquette1986,koester2009,heinonen2020}, a negligible timeframe compared to white dwarf evolutionary timescales. The detection of heavy elements in a white dwarf atmosphere can be naturally explained by the recent or ongoing accretion of rocky planetesimals \citep[e.g., see the reviews by][]{jura2014,farihi2016}, an hypothesis that is reinforced by the Earth-like abundance patterns measured in most of those ``polluted'' white dwarfs \citep{zuckerman2007,klein2010,gansicke2012,doyle2019,xu2019,harrison2021}. In addition, dozens of white dwarfs with dust or gas disks have been identified (either through their thermal infrared signature or through heavy element emission lines), and they all show evidence for planetary matter accretion in their atmospheres \citep{rocchetto2015,rebassa2019,wilson2019,manser2020}. A few white dwarfs exhibiting recurring transit events due to planetary debris have also been discovered \citep{vanderburg2015,vanderbosch2020,vanderbosch2021}, cementing the case for polluted white dwarfs being recent or ongoing accretors of rocky debris from their planetary systems.

Those accretion episodes imply the existence of mechanisms that can perturb planetesimals orbiting several AUs from their host white dwarfs and deliver them to the immediate vicinity of the star where they can be accreted following tidal disruption events. Explaining the origin of those perturbations, either through dynamical $N$-body simulations of planetary systems or through analytical work, is an area of active research, and many viable scenarios have been identified \citep{bonsor2011,debes2012,veras2013,veras2016,frewen2014,mustill2014,mustill2018,caiazzo2017,petrovich2017,stephan2017,smallwood2018,maldonado2020a,maldonado2020b,li2021}. For instance, a single Jovian planet interacting with an asteroid belt can produce the instabilities needed to deliver planetesimals inside the Roche radius of the white dwarf \citep{bonsor2011,debes2012,frewen2014}. A key prediction of several $N$-body simulations is that as the white dwarf ages, there are less and less planetesimals available to be scattered inwards and accreted by the central white dwarf, leading to a decrease in the frequency of accretion episodes over Gyr timescales \citep{bonsor2011,debes2012,veras2013,veras2016,frewen2014,mustill2014,antoniadou2016,li2021}. As instabilities are thought to be partially triggered by mass loss on the red and asymptotic giant branches, scattering events tend to happen at early cooling times rather than after several Gyrs on the white dwarf cooling track. However, the magnitude of the decrease of the frequency of accretion events varies greatly from one study to the other; it is particularly sensitive to the assumed planetary system architecture.

This predicted decrease can be tested empirically by studying the relation between the mass accretion rates of Gyr-old white dwarfs and their ages. A decrease in the accretion rate is indicative of a decrease in the number of large objects available for accretion and/or of a decrease in the frequency of tidal disruption events. This is a crucial test for dynamical models, and it also constitutes a unique way to probe the full architecture of extrasolar planetary systems since different objects are expected to be perturbed at different times \citep{li2021}. Recent studies of polluted white dwarf samples seem to support the existence of a strong decrease in accretion rates over Gyr timescales \citep{hollands2018,chen2019}, but earlier studies had found no conclusive evidence for such trend \citep{koester2006,koester2014,wyatt2014}.

New developments now allow us to reassess this question more accurately. In particular, previous works largely relied on samples where the masses of white dwarfs were unknown. The standard procedure is then to assume a $\log g=8$ surface gravity for all objects, which may bias the atmosphere and envelope modelling used to determine the accretion rates of those objects. This issue is now largely resolved thanks to Gaia DR2 \citep{gaiadr2a,gaiadr2b}. Gaia provides precise trigonometric parallaxes for the majority of spectroscopically identified white dwarfs, which in turn allows precise mass determinations \citep{bergeron1997,bergeron2019}. In addition, recent observational efforts aimed at securing spectra for new white dwarf candidates found thanks to Gaia \citep{gentile2019} have enabled the identification of a large number of new cool polluted white dwarfs \citep{kilic2020,tremblay2020}. This increase in the number of known polluted white dwarfs is crucial for cool H-atmosphere white dwarfs, for which very few polluted objects are known.

In this work, we investigate the relation between the inferred mass accretion rates \mdotz{} and ages of old white dwarfs (cooling time $\tau_{\rm cool}>1\,{\rm Gyr}$). We first present a brief digression in Section~\ref{sec:detection} to explain why cool polluted H-atmosphere white dwarfs are so rare compared to their He-atmosphere analogs. We then turn to our main goal of establishing the time evolution of the accretion rates of old white dwarfs in the subsequent sections. In Section~\ref{sec:samples}, we present our sample of 34 H- and 664 He-atmosphere objects, and perform a detailed atmospheric analysis of all our H-atmosphere objects (the He-atmosphere objects have been analysed elsewhere). We then investigate the relation between \mdotz{} and $\tau_{\rm cool}$ and discuss the implications of our findings in Section~\ref{sec:results}. Finally, we conclude in Section~\ref{sec:conclu}.

\section{On the scarcity of cool polluted H-atmosphere white dwarfs}
\label{sec:detection}
H-atmosphere white dwarfs are much more numerous than their He-atmosphere counterparts, representing $\simeq 80\%$ of the white dwarf population older than $\tau_{\rm cool}=1\,{\rm Gyr}$ \citep{blouin2019c}. Despite this fact, the population of old polluted white dwarfs is overwhelmingly dominated by He-atmosphere objects. This can be explained by detection limit considerations. Cool He atmospheres are much more transparent than their H counterparts, implying that a smaller amount of heavy elements is required for accretion to leave a detectable spectroscopic signature. 

To illustrate this point, we have determined the detection limit of Ca for all white dwarfs of the cool sample of \cite{blouin2019c} for which spectra are publicly available in the Montreal White Dwarf Database \citep[MWDD,][]{dufour2017}\footnote{\url{https://www.montrealwhitedwarfdatabase.org/}}. 
We use this particular sample for this exercise because (1) it is the largest sample of cool white dwarfs for which the atmospheric composition (H or He dominated) was accurately determined individually for each object and (2) it is effectively a magnitude-limited sample in $J$ and should be minimally biased against any spectral type \citep[see discussion in Section~4.1 of][]{blouin2019c}. Note that this sample is different from the one we use in the rest of this work to infer how \mdotz{} varies with \taucool{}. In the present case, we want a sample that is unbiased against any spectral type, while in the latter case, it is quite the opposite and we want to include as many polluted white dwarfs as we can.

\begin{figure}
	\includegraphics[width=\columnwidth]{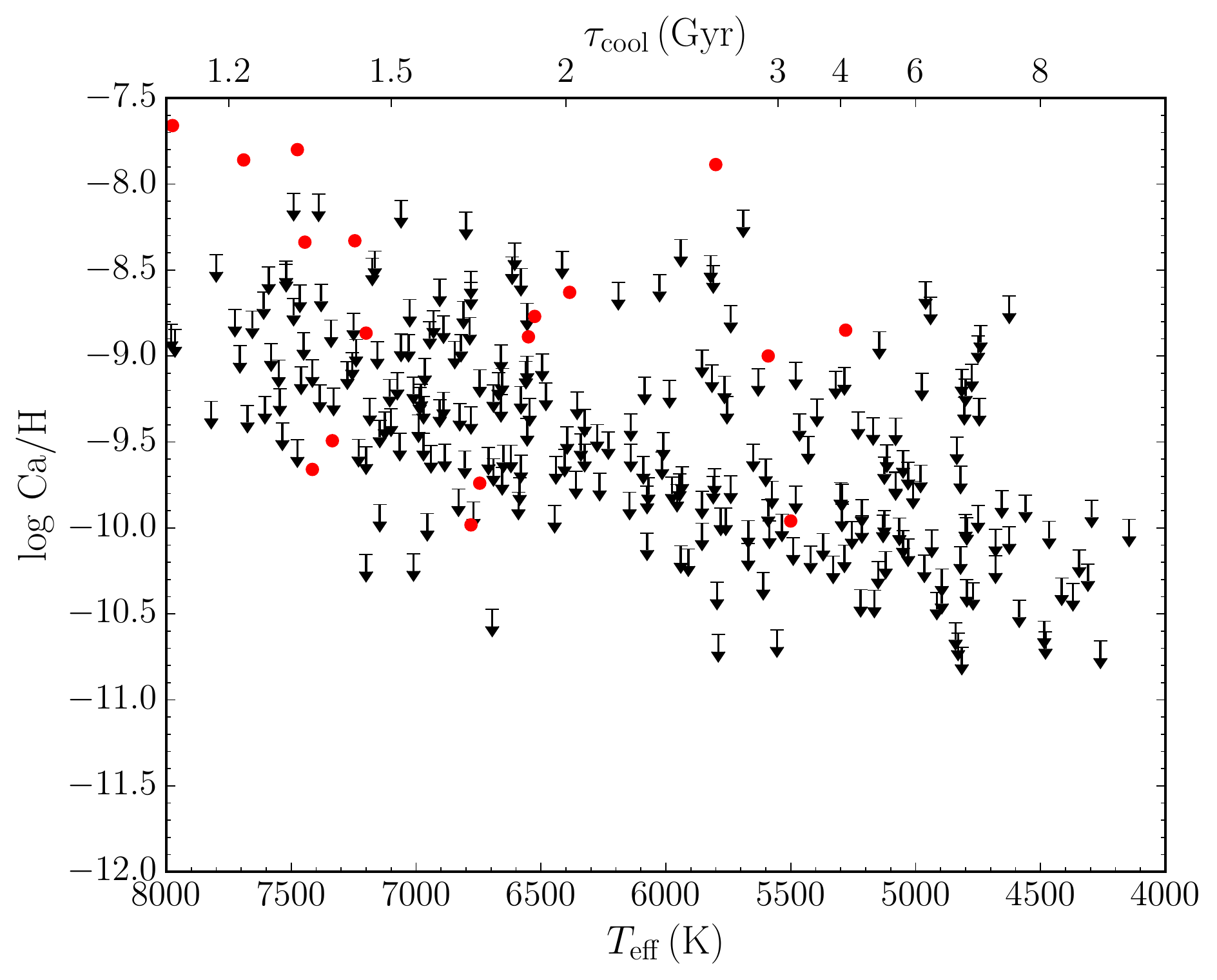}
	\includegraphics[width=\columnwidth]{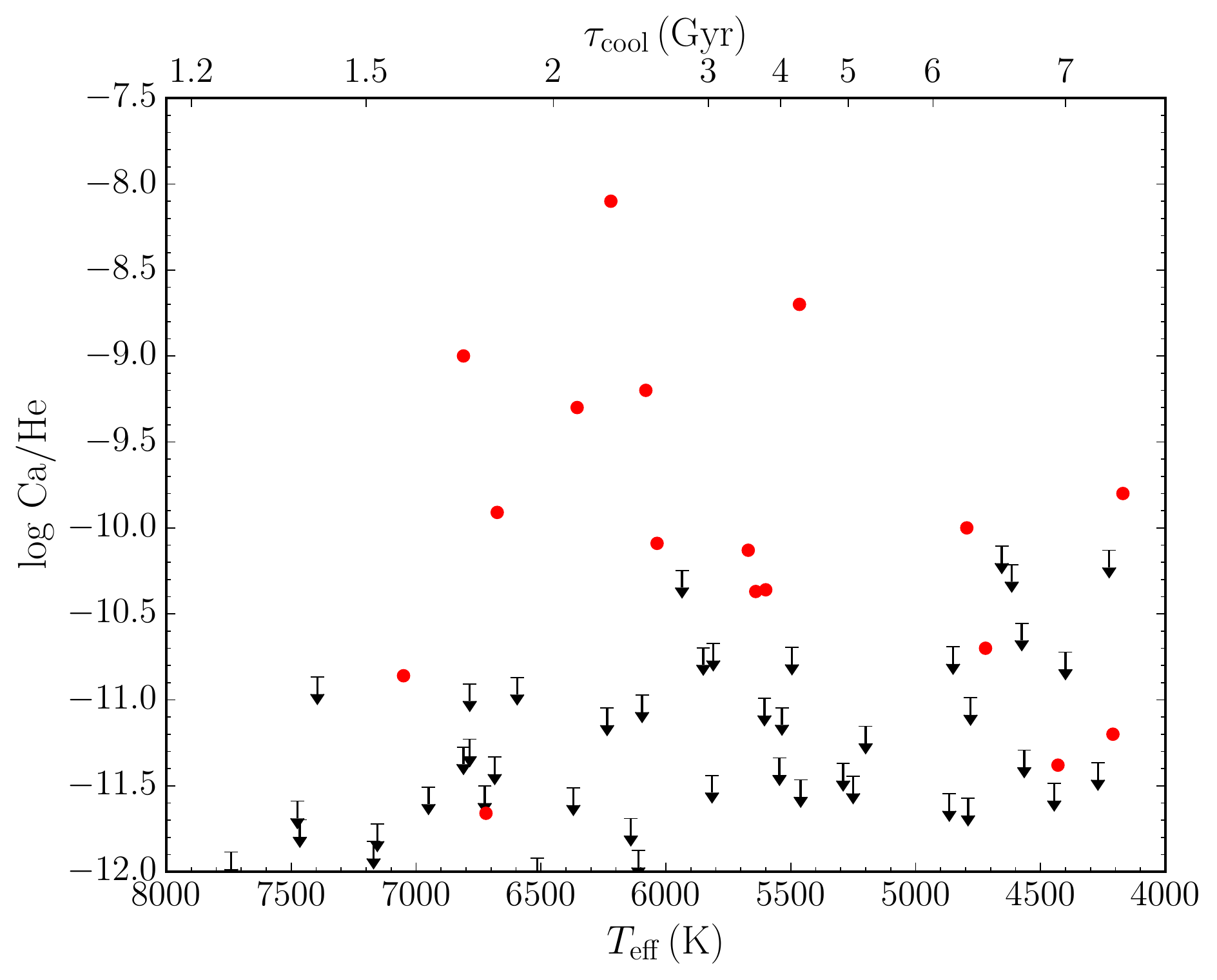}
    \caption{Ca abundances (red circles) and upper limits (arrows) for cool H- (top panel) and He-atmosphere (bottom panel) white dwarfs in the sample of \protect\cite{blouin2019c}. A given pollution level $\log\,{\rm Ca/H(e)}$ is more likely to be detected in an He- than in an H-atmosphere white dwarf. The upper horizontal axis gives the white dwarf cooling age assuming a mass of $0.6\,M_{\odot}$ and an H envelope of $M_{\rm H}/M_{\star}=10^{-4}$ (top panel) or $M_{\rm H}/M_{\star}=10^{-10}$ (bottom panel). Note that the atmospheric compositions can be uncertain for objects without Ca detections below $T_{\rm eff} \simeq 5000\,{\rm K}$ (when H$\alpha$ is no longer visible). The individual Ca upper limits at low temperatures should therefore be interpreted with caution, although the general trends depicted here and discussed in the text are robust.}
    \label{fig:Calimits}
\end{figure}

The results of this exercise are shown in Figure~\ref{fig:Calimits}, where red circles correspond to Ca detections and arrows indicate upper limits. Clearly, a given pollution level Ca/H(e)\footnote{All abundance ratios given in this work correspond ratios of number abundances.} is more likely to be detected in an He than in an H atmosphere, thereby explaining the paucity of cool polluted H-atmosphere white dwarfs. The detection limits of Ca in the He-atmosphere objects are clustered around $\log\,{\rm Ca/He}=-11$ and essentially independent of the white dwarf temperature, while there is a strong temperature dependence for H-atmosphere objects with detection limits of the order of $\log\,{\rm Ca/H}=-8.5$ at $T_{\rm eff}=8000\,{\rm K}$ to $\log\,{\rm Ca/H}=-10$ at $T_{\rm eff}=5000\,{\rm K}$.

\section{Samples and Atmospheric Parameters Determination}
\label{sec:samples}
\subsection{H-atmosphere white dwarfs}
The limited number of old polluted H-atmosphere white dwarfs poses a challenge to the study of the evolution of the accretion rate of those objects. \cite{chen2019} recently investigated the relation between \mdotz{} and \taucool{} for H-atmosphere white dwarfs compiled in the MWDD from a number of different studies. In total, their sample contains only 14~objects older than 1\,Gyr, a rather suboptimal situation to study the evolution of accretion rates over Gyr timescales. To increase this number, we selected all H-atmosphere white dwarfs in the MWDD that (1) were previously classified as cooler than $T_{\rm eff}=8000\,{\rm K}$ (which corresponds to $\tau_{\rm cool} \simeq 1\,{\rm Gyr}$ assuming a $0.6\,M_{\odot}$ mass), and (2) have a publicly available spectrum that clearly reveals the presence of heavy element pollution. We were able to identify 34~objects satisfying those criteria (Table~\ref{tab:daz}). The more than 2x increase in sample size compared to \cite{chen2019} is due in part to the recent identification of new polluted H-atmosphere white dwarfs \citep[e.g.,][]{kilic2020,tremblay2020}, but also to the fact that the heavy element abundances $\log\,{\rm Ca/H}$ of many previously identified DAZ white dwarfs had never been measured despite the availability of archival spectroscopy.

We determined the atmospheric parameters of our sample of H-atmosphere white dwarfs using an hybrid photometric/spectroscopic method \citep[e.g.,][]{dufour2007} and the atmosphere code described in \citet[and references therein]{blouin2018a,blouin2018b}. Synthetic photometry is first adjusted to Pan--STARRS {\it grizy} photometry \citep{chambers2016}, which yields a first estimate of $T_{\rm eff}$ and the solid angle $\pi R^2/D^2$. The Gaia parallax is then used to compute the radius $R$ of the white dwarf, from which we infer the mass (and $\log g$) using the evolutionary models of \cite{bedard2020}.\footnote{\url{https://www.astro.umontreal.ca/~bergeron/CoolingModels/}} This is done assuming a ``thick'' H envelope ($M_{\rm H}/M_{\star}=10^{-4}$), a standard assumption for H-atmosphere white dwarfs. Keeping $T_{\rm eff}$ and $\log g$ constant, we then turn to the spectroscopic data and adjust the Ca abundance to fit the \caii{} H and K lines. The photometric and spectroscopic fits are repeated until the atmospheric parameters converge to stable values, which occurs after only a few iterations. Note that the {\it grizy} photometry is dereddened using the same procedure as \cite{gentile2019} and the extinction maps of \cite{schlafly2011}.

\begin{figure*}
	\includegraphics[width=2\columnwidth]{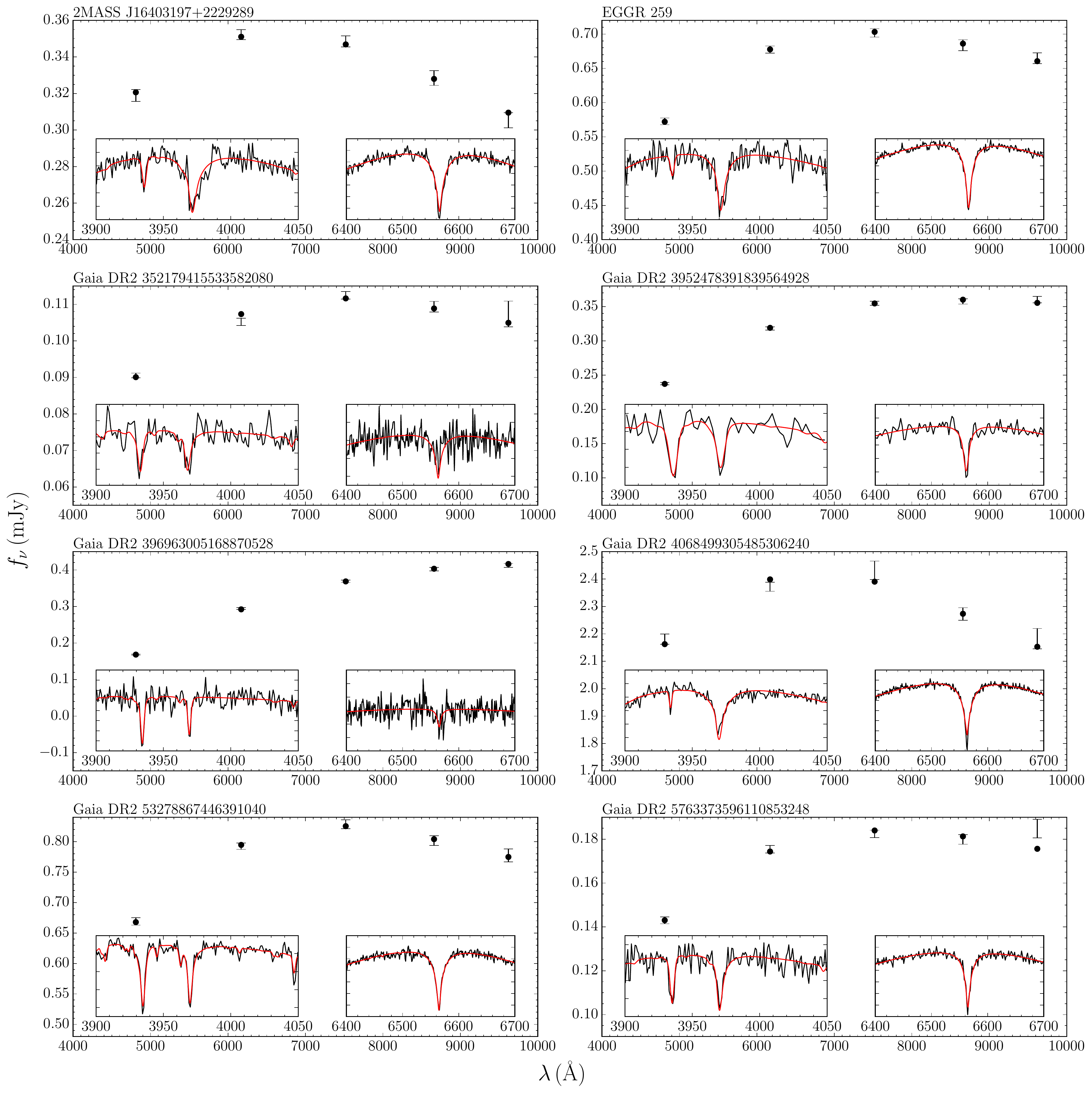}
    \caption{Photometric and spectroscopic fits for our H-atmosphere sample. The error bars represent the observed Pan--STARRS {\it grizy} photometry and the black circles the synthetic photometry of our best-fit models. For each panel, the left inset shows our fit to the \caii{} H and K lines, and the right inset displays the H$\alpha$ profile (or H$\beta$ when the spectral coverage does not extend to H$\alpha$) predicted by the photometric solution. The complete figure set is available in the online version of the journal.}
    \label{fig:fits}
\end{figure*}

Our best-fit parameters are given in Table~\ref{tab:daz}, where we also give the source of the spectra used in our atmospheric modelling. All spectra are publicly available on the MWDD website and the atmospheric parameters derived in this work will be incorporated to the MWDD. The uncertainties given in Table~\ref{tab:daz} correspond to the formal 1$\sigma$ uncertainties from the Levenberg--Marquardt algorithm. Note that the uncertainties on $\log\,{\rm Ca/H}$ include both the formal uncertainty on $\log\,{\rm Ca/H}$ at the nominal photometric $T_{\rm eff}$ and $\log g$ and that induced by the uncertainty on $T_{\rm eff}$ and $\log g$ \citep[see Equation~A1 of][]{klein2021}. Figure~\ref{fig:fits} displays our photometric and spectroscopic fits. Note that the H Balmer lines were not considered in our fits as we opted to determine $T_{\rm eff}$ and $\log g$ using the photometric technique (which allows us to take full advantage of Gaia parallaxes). Figure~\ref{fig:fits} shows the H$\alpha$ profiles predicted by our photometric solutions as a means to validate the photometric solutions and confirm the H-rich nature of those atmospheres.

\begin{table*}
	\centering
	\caption{Atmospheric parameters and inferred properties of polluted H-atmosphere white dwarfs analysed in this work.}
	\label{tab:daz}
	\begin{tabular}{lcccccccc} 
		\hline
		MWDD ID & Spectral & $T_{\rm eff}$ & $\log g$ & $M_{\star}$ & $\log\,{\rm Ca/H}^a$ & \taucool{} & $\log \dot{M}_Z$ & Spectrum\\
		 &  type & (K) & ($\log {\rm cm\,s}^{-2}$) & $(M_{\odot})$ &  & (Gyr) & ($\log {\rm g\,s}^{-1}$) & source \\
		\hline
		2MASS J16403197+2229289 &DAZ &$7560 \pm 55$ &$8.071 \pm 0.019$ &$0.637 \pm 0.012$ &$-8.33 \pm 0.08$ & 1.4 & 8.6 & 1 \\
EGGR 259 &DAZ &$6820 \pm 45$ &$7.993 \pm 0.017$ &$0.587 \pm 0.010$ &$-9.74 \pm 0.10$ & 1.7	&	7.2  & 1\\
Gaia DR2 352179415533582080 &DAZ &$6775 \pm 45$ &$8.175 \pm 0.036$ &$0.700 \pm 0.023$ &$-7.91 \pm 0.08$ & 2.3	&	9.2 &2 \\
Gaia DR2 3952478391839564928 &DAZ &$6025 \pm 25$ &$8.032 \pm 0.015$ &$0.607 \pm 0.009$ &$-8.59 \pm 0.07$ & 2.5	&	8.6 & 3 \\
Gaia DR2 396963005168870528 &DZA &$5005 \pm 15$ &$7.998 \pm 0.014$ &$0.578 \pm 0.009$ &$-9.63 \pm 0.07$ & 5.9	&	7.9 &4 \\
Gaia DR2 4068499305485306240 &DAZ &$7425 \pm 65$ &$8.061 \pm 0.019$ &$0.630 \pm 0.012$ &$-9.34 \pm 0.09$ & 1.5	&	7.6 & 4 \\
Gaia DR2 53278867446391040 &DAZ &$6785 \pm 45$ &$8.223 \pm 0.015$ &$0.732 \pm 0.010$ &$-7.65 \pm 0.06$ & 2.7	&	9.5 &4 \\
Gaia DR2 5763373596110853248 &DAZ &$6605 \pm 50$ &$7.984 \pm 0.028$ &$0.581 \pm 0.017$ &$-8.86 \pm 0.07$ & 1.8	&	8.2 & 1 \\
Gaia DR2 637527525030988160 &DAZ &$6240 \pm 45$ &$8.159 \pm 0.025$ &$0.688 \pm 0.016$ &$-8.14 \pm 0.07$ & 3.0	&	9.1 & 2 \\
Gaia DR2 657673498630210176 &DAZ &$6350 \pm 45$ &$8.123 \pm 0.036$ &$0.665 \pm 0.023$ &$-8.82 \pm 0.08$ & 2.5	&	8.3 & 1 \\
Gaia DR2 866962856917166208 &DAZ &$5740 \pm 35$ &$8.062 \pm 0.023$ &$0.624 \pm 0.014$ &$-8.78 \pm 0.07$ & 3.1	&	8.5 & 1 \\
GJ 1042 &DAZ & $7225 \pm 65$&	$7.970 \pm 0.020$&	$0.574 \pm 0.012$&	$-8.86 \pm 0.07$& 1.4	&	8.0 & 8\\
GJ 1052 &DAZ & $6785 \pm 40$&	$8.156 \pm 0.014$&	$0.688 \pm 0.009$&	$-9.95 \pm 0.08$& 2.3	&	7.1 & 8\\
LP 20$-$214$^b$ &DZ &$4910 \pm 40$ &$7.925 \pm 0.026$ &$0.534 \pm 0.015$ &$-9.13 \pm 0.07$ & 5.5	&	8.4 &3 \\
LP 815$-$31 &DAZ &$5505 \pm 20$ &$7.926 \pm 0.012$ &$0.541 \pm 0.007$ &$-9.97 \pm 0.07$ & 2.7	&	7.3 & 5 \\
LSPM J0132+0529 &DAZ &$7370 \pm 60$ &$7.968 \pm 0.031$ &$0.574 \pm 0.018$ &$-6.89 \pm 0.06$ & 1.3	&	10.0 & 1 \\
LSPM J0543+3637 &DAZ &$6410 \pm 35$ &$8.048 \pm 0.014$ &$0.619 \pm 0.009$ &$-8.63 \pm 0.06$ & 2.1	&	8.5 &3 \\
SDSS J010749.34+210745.2 &DAZ &$7695 \pm 195$ &$8.742 \pm 0.054$ &$1.066 \pm 0.031$ &$-7.50 \pm 0.18$ & 4.0	&	9.7 &1 \\
SDSS J082611.69+325000.1 &DAZ &$6005 \pm 30$ &$8.000 \pm 0.021$ &$0.588 \pm 0.013$ &$-9.21 \pm 0.08$ & 2.3	&	8.0 &1 \\
SDSS J094714.36+455701.3 &DAZ &$7610 \pm 80$ &$8.074 \pm 0.049$ &$0.638 \pm 0.030$ &$-5.69 \pm 0.08$ & 1.4	&	11.2 &1 \\
SDSS J102542.85+335410.9 &DAZ &$7520 \pm 100$ &$7.792 \pm 0.152$ &$0.477 \pm 0.078$ &$-7.50 \pm 0.16$ & 1.0	&	9.2 & 1 \\
SDSS J102844.04+162638.9 &DAZ &$8000 \pm 150$ &$8.075 \pm 0.279$ &$0.640 \pm 0.169$ &$-7.55 \pm 0.22$ & 1.2	&	9.3 &1 \\
SDSS J110318.87+393558.3 &DAZ &$4995 \pm 30$ &$7.910 \pm 0.021$ &$0.526 \pm 0.012$ &$-9.17 \pm 0.07$ & 4.7	&	8.3 &1 \\
SDSS J134856.00+364617.5 &DAZ &$7655 \pm 105$ &$8.125 \pm 0.121$ &$0.671 \pm 0.076$ &$-8.17 \pm 0.16$ & 1.5	&	8.7 &1 \\
SDSS J142242.42+593802.0 &DAZ &$7520 \pm 190$ &$8.191 \pm 0.090$ &$0.712 \pm 0.058$ &$-7.85 \pm 0.19$ & 1.8	&	9.1 &1 \\
SDSS J164539.75+305931.6 &DAZ &$7080 \pm 55$ &$8.167 \pm 0.023$ &$0.695 \pm 0.015$ &$-7.37 \pm 0.08$ & 2.0	&	9.7 &1 \\
Ton 953 &DAZ &$7450 \pm 60$	&$7.954 \pm 0.018$&	$0.566 \pm 0.010$&	$-9.66 \pm 0.09$& 1.2	&	7.2 &1\\
WD 0850+190 &DAZ &$8085 \pm 70$ &$8.004 \pm 0.026$ &$0.597 \pm 0.016$ &$-6.99 \pm 0.07$ & 1.1	&	9.8& 1 \\
WD 0920+012 &DAZ &$6070 \pm 25$ &$7.981 \pm 0.013$ &$0.577 \pm 0.008$ &$-8.93 \pm 0.08$ & 2.2	&	8.2 &6 \\
WD 1408+029 &DAZ &$5595 \pm 20$ &$7.991 \pm 0.013$ &$0.580 \pm 0.008$ &$-9.00 \pm 0.07$ & 2.9	&	8.3 &6 \\
WD 1626+490 &DZA &$5255 \pm 30$ &$8.101 \pm 0.021$ &$0.645 \pm 0.014$ &$-9.00 \pm 0.07$ & 5.6	&	8.5 &2 \\
WD 1653+385 &DAZ &$5790 \pm 30$ &$8.178 \pm 0.014$ &$0.698 \pm 0.009$ &$-7.88 \pm 0.07$ & 4.1	&	9.5 &7 \\
WD 2152+127 &DZA &$4930 \pm 30$ &$7.838 \pm 0.025$ &$0.485 \pm 0.013$ &$-9.74 \pm 0.08$ & 4.3	&	7.7 &2 \\
WD 2225+176 &DAZ &$6605 \pm 40$ &$7.943 \pm 0.016$ &$0.557 \pm 0.009$ &$-8.89 \pm 0.08$ & 1.7	&	8.1 &3 \\
		\hline
		\multicolumn{9}{l}{\parbox{2\columnwidth}{Spectrum source: (1) SDSS; (2) \cite{kilic2020}; (3) \cite{limoges2015}; (4) \cite{tremblay2020}; (5) \cite{subasavage2017}; (6) \cite{sayres2012}; (7) \cite{giammichele2012}; (8) \cite{gianninas2011}}}\\
		\multicolumn{9}{l}{\parbox{2\columnwidth}{$^a$ Number abundances}}\\
		\multicolumn{9}{l}{\parbox{2\columnwidth}{$^b$ This object is classified as DZ due to the non-detection of Balmer lines, but it has an H-dominated atmosphere based on the width of its \caii{} H and K lines (see \citealt{mccleery2020} and our fit in the extended version of Figure~\ref{fig:fits}).}}\\
		\end{tabular}
\end{table*}

\subsection{He-atmosphere white dwarfs}
Large samples of cool polluted He-atmosphere white dwarfs have already been analysed in previous studies \citep{dufour2007,koester2011,hollands2017,coutu2019,blouin2020}. With 664 polluted He-atmosphere white dwarfs cooler than $T_{\rm eff}=8000\,{\rm K}$, the sample of \cite{coutu2019} is the largest (it includes virtually all polluted He-atmosphere objects listed in the MWDD). Also, the atmospheric parameters reported in \cite{coutu2019} are based on the same state-of-the-art model atmosphere code \citep{blouin2018a,blouin2018b} used in the present work for our H-atmosphere sample. For those reasons, those 664 objects define our He-atmosphere sample. We directly use the atmospheric parameters derived in \cite{coutu2019}, but, when available, we replace them by those obtained by \cite{blouin2020}. In this last study, the same model atmosphere code was used, but the Ca, Fe, and Mg abundances were individually determined, whereas \cite{coutu2019} only fitted the Ca abundance. This can induce a systematic bias in the solutions, so we rely on the more detailed analyses of \cite{blouin2020} for the 193 objects where they are available.

\section{Results and Discussion}
\label{sec:results}
\subsection{Accretion rates calculation}
Under the assumption of a steady-state equilibrium between accretion onto the white dwarf and diffusion at the bottom of the convection zone, the mass accretion rate can be computed as \citep[e.g.,][]{koester2009}
\begin{equation}
    \dot{M}_{\scriptscriptstyle Z} = \frac{1}{A}\frac{X_{\rm Ca} M_{\rm cvz}}{\tau_{\rm Ca}},
    \label{eq:Mdot}
\end{equation}
where $A$ is the mass fraction of Ca in the accreted material (which, following \citealt{farihi2012}, we assume to be the value for the bulk Earth [0.016], \citealt{allegre1995}), $X_{\rm Ca}$ is the mass fraction of Ca in the envelope (as inferred from the photospheric Ca abundance), $M_{\rm cvz}$ is the mass of the convective zone, and \tauca{} is the diffusion timescale of Ca at the bottom of the convection zone. We use $M_{\rm cvz}$ and $\tau_{\rm Ca}$ values taken from the envelope calculations of \cite{koester2020} to evaluate \mdotz{} by virtue of Equation~\ref{eq:Mdot}. Those $M_{\rm cvz}$ and $\tau_{\rm Ca}$ tables are available online\footnote{\url{http://www1.astrophysik.uni-kiel.de/~koester/astrophysics/astrophysics.html}}, but for He-atmosphere white dwarfs we use updated tables provided by D.~Koester (private communication). Those tables are based on envelope calculations that use polluted (instead of pure H/He) atmosphere models as boundary conditions \citep[e.g., see][]{harrison2021}.

It is important to note that \tauca{} is large ($10^3-10^7$ years) for our sample of cool objects, meaning that a balance between accretion and diffusion is probably not achieved for all objects. The accretion rates reported below should therefore be viewed as average rates over $\tau_{\rm Ca}$ and not as instantaneous rates \citep{farihi2012}. This average accretion rate can differ significantly from the instantaneous rate if accretion is dominated by sporadic high-rate episodes \citep{girven2012,farihi2012,xu2016}.

\subsection{H-atmosphere white dwarfs}
\begin{figure}
	\includegraphics[width=\columnwidth]{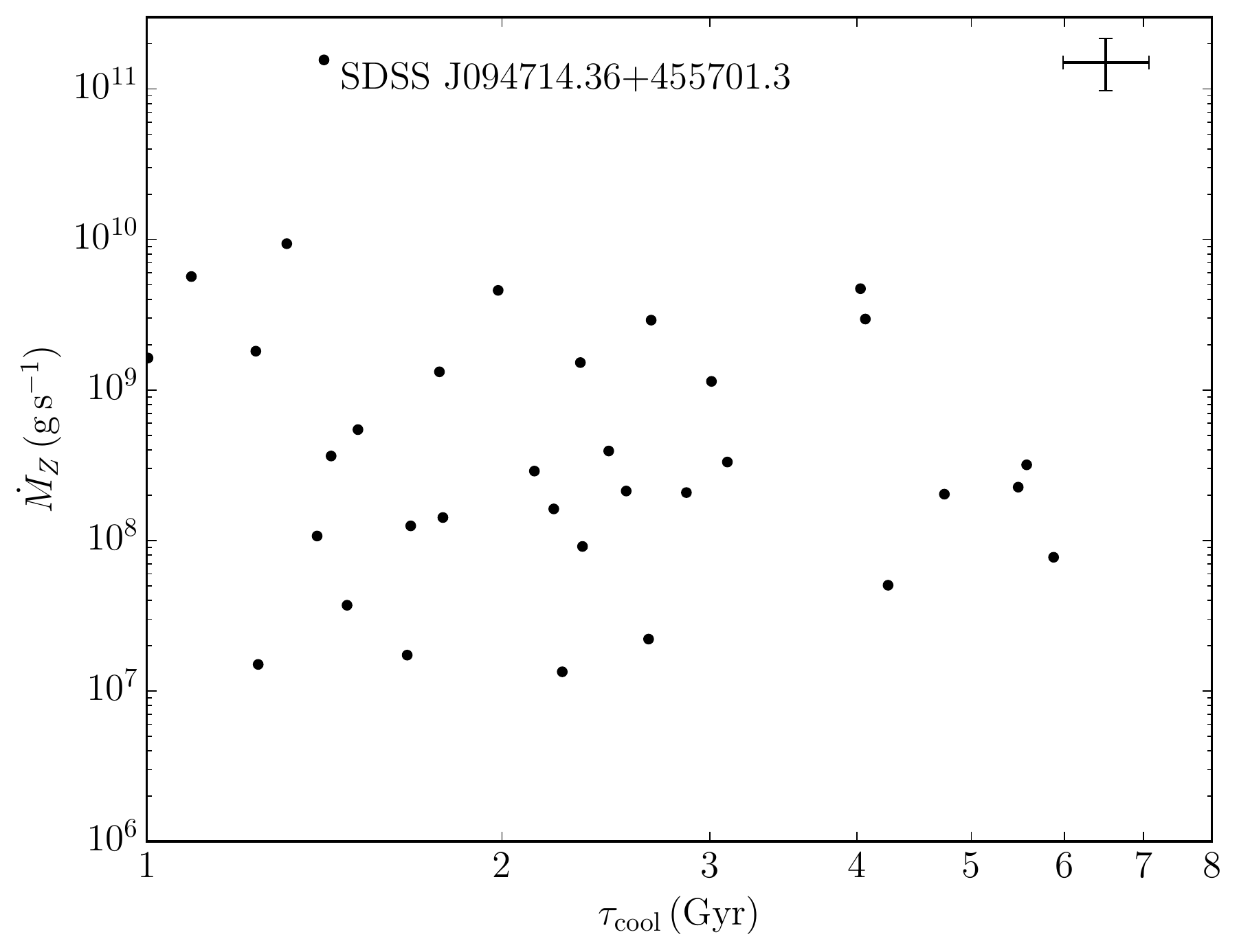}
    \caption{Mass accretion rate as a function of white dwarf cooling time for the polluted H-atmosphere white dwarf sample analysed in this work (Table~\ref{tab:daz}). The error bars illustrate the typical uncertainties on each data point. All data points correspond to Ca detections. The upper envelope of the distribution is poorly defined, which prevents the identification of any clear trend between \mdotz{} and \taucool{}. However, it does not support the strong decay implied by the power law of \citealt{chen2019} (see text for details).}
    \label{fig:Mdot_blouin}
\end{figure}

Figure~\ref{fig:Mdot_blouin} shows the evolution of \mdotz{} for our H-atmosphere sample. The first thing to note is that only the upper envelope of the distribution (i.e., the most polluted objects) has a physical meaning. The low-\mdotz{} edge depends on visibility limits considerations. It is affected by variables that have nothing to do with accretion itself, such as the signal-to-noise ratio of the spectra that happen to be available for cool H-atmosphere white dwarfs and the general trend that exists between the Ca detection limit and the effective temperature (Figure~\ref{fig:Calimits}).

Even though our cool polluted H-atmosphere sample is significantly larger than those considered in previous studies, the upper envelope of \mdotz{} vs \taucool{} remains poorly defined and we are unable to discern any clear trend. Nevertheless, Figure~\ref{fig:Mdot_blouin} appears to be at odds with the strong decay rate implied by the power-law fit of \cite{chen2019} for old H-atmosphere white dwarfs,
\begin{equation}
\log \left( \dot{M}_{\scriptscriptstyle Z} / {\rm g\,s}^{-1} \right) = -2.07 \times \log \left( \tau_{\rm cool} / {\rm Gyr} \right) + {\rm constant}.
\label{eq:chen}
\end{equation}
Leaving the $\dot{M}_{\scriptscriptstyle Z} \simeq 10^{11}\,{\rm g\,s}^{-1}$ outlier (SDSS~J094714.36+455701.3) aside\footnote{SDSS~J094714.36+455701.3 was first identified by \cite{kleinman2013}, but its Ca abundance had never been determined before. Its high pollution level makes it one of the most rapidly accreting polluted white dwarfs ever identified. Its SDSS spectrum also shows Fe and Mg lines that are consistent with our high Ca/H value.}, the upper envelope of Figure~\ref{fig:Mdot_blouin} does not offer strong support for a 1~dex decrease in \mdotz{} between 1 and 3~Gyr as implied by this power law. A weaker decay is however entirely compatible with our data, but we refrain from quantifying this decay given the poor definition of the upper envelope of \mdotz{} vs \taucool.

Note that \cite{chen2019} obtained this power-law trend (Equation~\ref{eq:chen}) by fitting all their data points, not just the upper envelope. They justify this approach by arguing that (1) the Ca detection limit is almost independent of \taucool{} for old white dwarfs (implying minimal observational bias) and that (2) the accretion decay can also be clearly seen from the upper envelope of the data. We doubt the validity of those two arguments. First, we can conclude from Figure~\ref{fig:Calimits} that the Ca detection limit varies by $\simeq 1.5\,{\rm dex}$ between $T_{\rm eff}=8000$ and 5000\,K. This implies a detection limit (both in terms of photospheric $\log {\rm Ca/H}$ and of inferred \mdotz{}) that is not flat with respect to \taucool{}. In addition, their sample was collated from many different sources, meaning that there is no single detection limit for a given temperature. The resolution and signal-to-noise ratio of the spectra of known DAZ white dwarfs vary greatly, and this non-uniformity can induce false trends in this small sample. For example, the power law of \cite{chen2019} is driven in part by the very low Ca pollution level of the old DAZ G174--14 \citep[$\log {\rm Ca/H}=-12.7$, $T_{\rm eff}=5190$K,][]{zuckerman2003}. We know this object is polluted only because it happened to have been observed using Keck/HIRES (an SDSS spectrum for instance would likely miss the Ca lines). This illustrates how the lower envelope of \mdotz{} vs \taucool{} can be influenced by considerations that have nothing to do with the physical mechanisms behind accretion. Secondly, given the size of their sample, an upper envelope cannot be robustly identified at large cooling times, much less the trend of this upper envelope. In short, we do not see evidence for a strong decay of the accretion rate at large cooling times in our data, and we do not see evidence for such decay in the smaller sample of \cite{chen2019} either. 

Thermohaline mixing is expected to take place in the envelopes of accreting white dwarfs \citep{deal2013,wachlin2017,bauer2018}. The mean molecular weight of the accreted material is higher than that of the envelope, which corresponds to an unstable configuration that induces extra mixing of the heavy elements. This effect is not accounted for in accretion rates inferred using Equation~\eqref{eq:Mdot}. \cite{wachlin2021} recently published estimates of accretion rates that include the extra mixing resulting from thermohaline convection. Using their Table~3, we verified that once thermohaline mixing is included our conclusions remain qualitatively unchanged (i.e., the upper envelope of \mdotz{} as a function of \taucool{} is not compatible with the strong decay rate of \citealt{chen2019}). However, it should be noted that \cite{wachlin2021} did not model objects cooler than $T_{\rm eff}=6000\,{\rm K}$ (which prevents us from extending this analysis to the oldest objects of our sample), and they only modelled $0.609\,M_{\odot}$ white dwarfs (which forced us to assume a constant mass for this analysis).

\subsection{He-atmosphere white dwarfs}
\begin{figure}
	\includegraphics[width=\columnwidth]{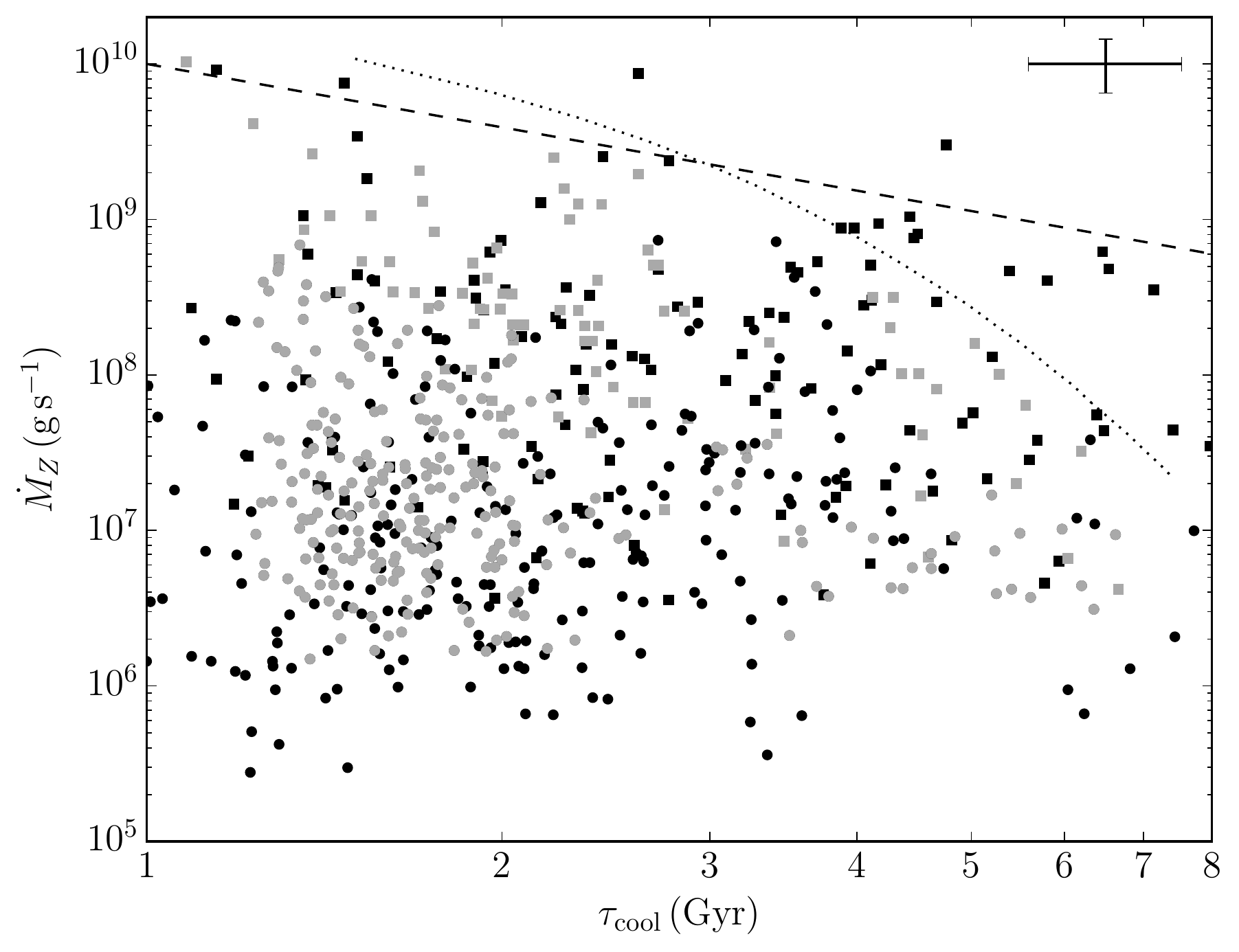}
    \caption{Mass accretion rate as a function of white dwarf cooling time for objects cooler than $T_{\rm eff}=8000\,{\rm K}$ in the \protect\cite{coutu2019} DZ sample. The subset of objects that were re-analysed in more details in \protect\cite{blouin2020} are shown as squares. The error bars illustrate the typical uncertainties on each data point. The dotted line shows the exponential trend identified in \protect\cite{hollands2018}. Black symbols represent white dwarfs for which a mass was secured from a Gaia parallax; grey symbols are for objects where $\log g=8$ was assumed. All data points correspond to Ca detections. The upper envelope of the \mdotz{} distribution reveals a tentative $\simeq 1\,{\rm dex}$ decrease between $\tau_{\rm cool}=1\,{\rm Gyr}$ and 8\,Gyr (dashed line); this is a much weaker decay than what was found in recent studies \protect\citep{hollands2018}.}
    \label{fig:Mdot_coutu}
\end{figure}

We now turn to our He-atmosphere sample. Being more than one order of magnitude larger than our H-atmosphere sample, it can provide more stringent constraints on the evolution of the accretion rate (Figure~\ref{fig:Mdot_coutu}). This time, the upper envelope is better defined and we can tentatively identify a weak decay of $\simeq 1\,{\rm dex}$ in \mdotz{} between $\tau_{\rm cool}=1$ and 8\,Gyr (dashed line), which would also be compatible with our results for H-atmosphere white dwarfs shown in Figure~\ref{fig:Mdot_blouin}. This is to be contrasted with the results of \cite{hollands2018}, who found an exponential decay with an e-folding timescale of $\simeq 1\,{\rm Gyr}$. This exponential trend (shown as a dotted line in Figure~\ref{fig:Mdot_coutu}) can be ruled out. 

The analysis of \cite{hollands2018} relied on a sample of 230 DZ white dwarfs for which a standard $\log g=8$ surface gravity was assumed \citep{hollands2017}. Interestingly, the exponential decay identified by \cite{hollands2018} provides a good match to the upper envelope of the subset of objects in the \cite{coutu2019} sample for which $\log g=8$ was assumed due to the absence of a Gaia parallax measurement (those objects are represented by grey symbols in Figure~\ref{fig:Mdot_coutu}). Only the objects for which a mass could be determined (black symbols) fill the gap above the exponential fit of \cite{hollands2018} and point toward a less steep decrease of \mdotz{}. This result suggests that accurate mass determinations are crucial for this type of study, and that all results obtained assuming a standard $\log g=8$ surface gravity should be interpreted with caution. 

\begin{figure}
	\includegraphics[width=\columnwidth]{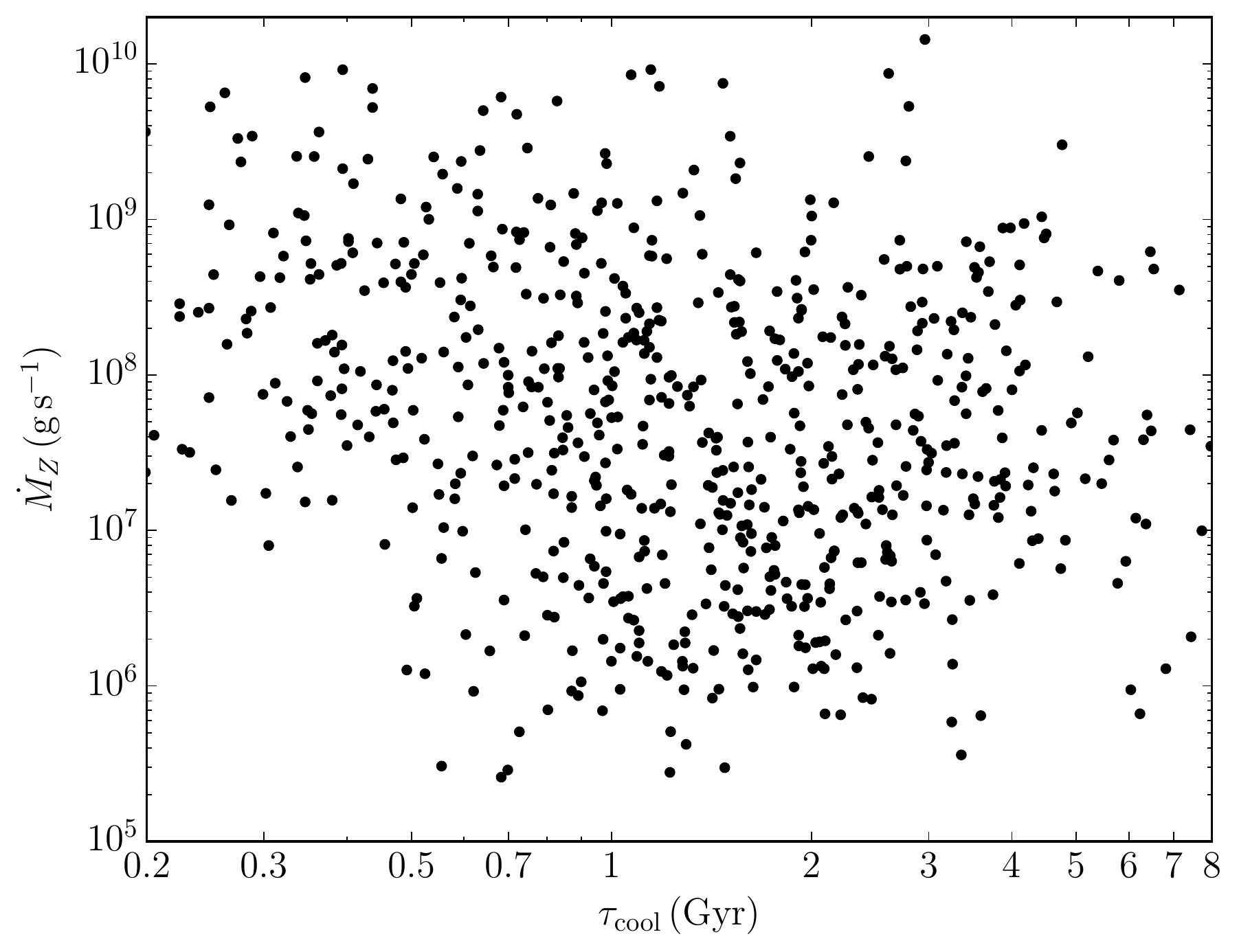}
    \caption{As Figure~\ref{fig:Mdot_coutu} but extended to objects of any effective temperature. This time, objects for which the mass is unknown are not shown. As in Figure~\ref{fig:Mdot_coutu}, the atmospheric parameters of \protect\cite{blouin2020} are used when available.}
    \label{fig:Mdot_coutu_hot}
\end{figure}

We extend our analysis of the \cite{coutu2019} sample to $\tau_{\rm cool}<1\,{\rm Gyr}$ in Figure~\ref{fig:Mdot_coutu_hot}. This time, we do not apply any temperature cut, but, based on our findings from the previous paragraph, we restrict the sample to objects with secure mass determinations only. This yields a sample of 685 objects. We see no evidence for a decrease of the upper envelope of \mdotz{} at $\tau_{\rm cool}<1\,{\rm Gyr}$; only in the $\tau_{\rm cool}=1-8\,{\rm Gyr}$ interval can a tentative $\simeq 1\,{\rm dex}$ decrease be identified. This is significantly weaker than what was found in recent studies.

\subsection{Comparison to the predictions of dynamical simulations}
Does the picture depicted above for H- and He-atmosphere white dwarfs (i.e., weakly decreasing accretion rates at 1--8~Gyr cooling times) disagree with the current state-of-the-art dynamical simulations? Unfortunately, we do not think this is a question that can be clearly answered at this time. The parameter space of possible planetesimal delivery mechanisms and planetary system architectures is extremely large \citep[e.g.,][]{brouwers2021,veras2021}. Each simulation has to make a set of assumptions and focus only on a small fraction of this parameter space, leading to different conclusions from one study to the other for the relation between \mdotz{} and \taucool{}. For instance, based on their simulations of a simple system consisting of a planet and a planetesimal belt orbiting a central star, \cite{bonsor2011} predict a weak decrease of \mdotz{} at large cooling times that is compatible with our empirical constraints (see their Figure~7). On the other hand, in their attempt to predict the future pollution of the solar white dwarf, \cite{li2021} find a decrease that is too strong compared to our results (see their Figure~9).

However, what we can confidently conclude from our data is that efficient mechanisms must still exist at large cooling times to deliver planetary materials to the immediate vicinity of white dwarfs. Recent results suggest that accretion events at late cooling times can be more easily achieved when instabilities are generated by terrestrial than by Jovian planets \citep{veras2021b}. The latter quickly eject asteroids and debris out of the system early in the evolution of the white dwarf, while smaller planets can continue to generate instabilities at larger cooling times.

\subsection{Comparison of H- and He-atmosphere white dwarfs}
Among younger ($\tau_{\rm cool} \lesssim 1\,{\rm Gyr}$) polluted white dwarfs, it is well established that He-atmosphere white dwarfs exhibit significantly higher time-averaged accretion rates than their H-atmosphere analogs \citep{farihi2012,girven2012,cunningham2021}. This is expected if accretion is dominated by short-lived\footnote{Those events have been estimated to last less than $10^3$ years \citep{farihi2012}.} high-rate episodes. At $T_{\rm eff}=15{,}000\,{\rm K}$, the diffusion timescale of Ca in an He atmosphere is of the order of $10^6$ years, while it is only of the order of a few days in an H atmosphere. Therefore, the signature of those sporadic high-rate episodes on the time-averaged accretion rate is much more likely to be detected in an He atmosphere than in an H atmosphere. The longer diffusion timescales of He-atmosphere objects confer them a longer memory of past high-rate episodes, while the short diffusion timescales of their H-atmosphere counterparts imply that their time-averaged accretion rates rapidly ``forget'' about those events.

\begin{figure}
	\includegraphics[width=\columnwidth]{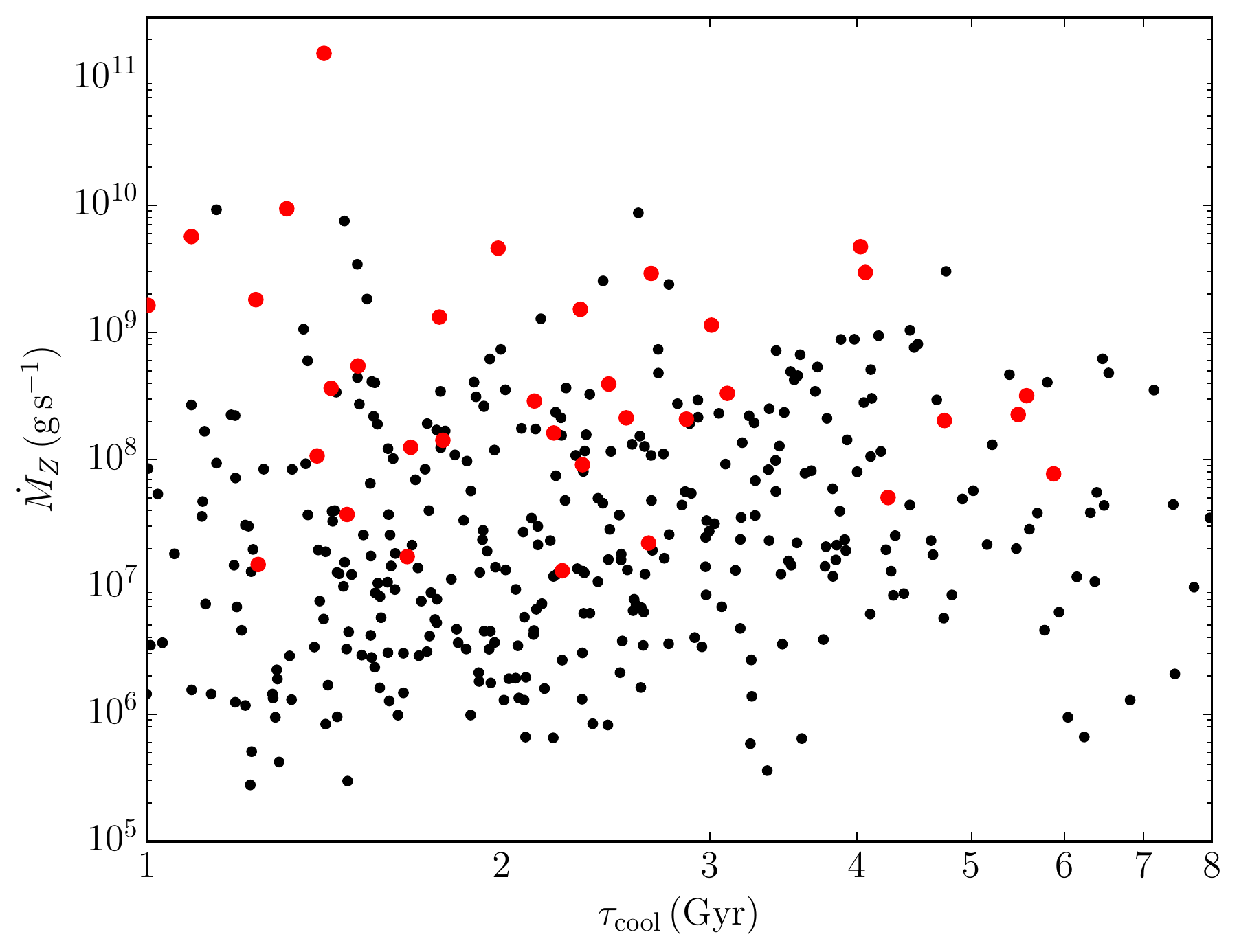}
    \caption{Comparison of the mass accretion rates for the H- (large red circles) and He-atmosphere (small black circles) samples. The data shown here is identical to Figure~\ref{fig:Mdot_blouin} and \ref{fig:Mdot_coutu}, except that objects for which the mass is unknown are not shown. The accretion rates of H- and He-atmosphere white dwarfs are very similar.}
    \label{fig:Mdot_comp}
\end{figure}

If this interpretation is correct, then one would expect a different picture for older, cooler white dwarfs. At $T_{\rm eff}=5000\,{\rm K}$, the diffusion timescales of H and He atmospheres is of the order of $10^5$ and $10^7$ years, respectively. The diffusion timescales being much more similar than for warm white dwarfs, the H and He atmospheres of older white dwarfs should have a similar memory of past high-rate events, and their time-averaged accretion rates should be more similar. Our results are in good agreement with this prediction. The upper envelopes of \mdotz{} for our samples of cool H- and He-atmosphere white dwarfs are very similar (Figure~\ref{fig:Mdot_comp}). We can firmly rule out the kind of difference observed for warmer white dwarfs, where the historical accretion rates of He-atmosphere white dwarfs can be $1-2$ orders of magnitude above the highest accretion rates inferred for H atmospheres \citep{girven2012}. Note that the good agreement between H- and He-atmosphere objects in Figure~\ref{fig:Mdot_comp} is also comforting with respect to the reliability of current envelope models. It suggests the absence of significant systematic errors between H and He envelope calculations that would affect the $M_{\rm cvz}$ or \tauca{} values.

\section{Summary and outlook}
\label{sec:conclu}
We have studied the relation between the time-averaged accretion rates and cooling ages of old polluted white dwarfs. To this end, we presented a detailed model atmosphere analysis of 34 cool polluted hydrogen-dominated white dwarfs, most of them being fitted for a metal abundance for the first time. At odds with recent studies, we found no compelling evidence for a strong decay of the inferred rocky material accretion rates for cooling ages greater than 1\,Gyr. Our results imply that accretion rates only decrease weakly over Gyr timescales. We have also shown that the accretion rates of old H- and He-atmosphere white dwarfs are very similar, a situation that differs significantly from previous findings for younger, warmer white dwarfs. This difference is naturally explained by the similarly large diffusion timescales of H and He atmospheres at low effective temperatures, which makes them equally able of maintaining a memory of past high-rate accretion events.

Our results indicate that rocky debris can still be efficiently delivered inside the Roche radius of white dwarfs even after several Gyrs on the cooling track. This is a strong constraint for the $N$-body dynamical simulations employed to explain accretion events, and we hope this will stimulate further modelling efforts to identify planetary systems architectures that can sustain high accretion rates over several Gyrs.

\section*{Acknowledgements}
We thank the referee, D.~Koester, for providing helpful comments that have improved this manuscript and for sharing with us updated diffusion timescales tables for He-atmosphere white dwarfs. We also thank D.~Veras for illuminating exchanges on dynamical simulations of evolved planetary systems, and we are grateful to the many researchers (including M.~Kilic, P.~Bergeron, J.~P. Subasavage, and P.-E. Tremblay) who generously made their white dwarf spectra publicly available on the MWDD. SB is a Banting Postdoctoral Fellow and a CITA National Fellow, supported by the Natural Sciences and Engineering Research Council of Canada (NSERC).

This work has made use of data from the European Space Agency (ESA) mission {\it Gaia} (\url{https://www.cosmos.esa.int/gaia}), processed by the {\it Gaia} Data Processing and Analysis Consortium (DPAC, \url{https://www.cosmos.esa.int/web/gaia/dpac/consortium}). Funding for the DPAC has been provided by national institutions, in particular the institutions participating in the {\it Gaia} Multilateral Agreement.

The Pan--STARRS1 Surveys (PS1) and the PS1 public science archive have been made possible through contributions by the Institute for Astronomy, the University of Hawaii, the Pan-STARRS Project Office, the Max-Planck Society and its participating institutes, the Max Planck Institute for Astronomy, Heidelberg and the Max Planck Institute for Extraterrestrial Physics, Garching, The Johns Hopkins University, Durham University, the University of Edinburgh, the Queen's University Belfast, the Harvard-Smithsonian Center for Astrophysics, the Las Cumbres Observatory Global Telescope Network Incorporated, the National Central University of Taiwan, the Space Telescope Science Institute, the National Aeronautics and Space Administration under Grant No. NNX08AR22G issued through the Planetary Science Division of the NASA Science Mission Directorate, the National Science Foundation Grant No. AST--1238877, the University of Maryland, Eotvos Lorand University (ELTE), the Los Alamos National Laboratory, and the Gordon and Betty Moore Foundation.

This work is partly supported by the international Gemini Observatory, a program of NSF's NOIRLab, which is managed by the Association of Universities for Research in Astronomy (AURA) under a cooperative agreement with the National Science Foundation, on behalf of the Gemini partnership of Argentina, Brazil, Canada, Chile, the Republic of Korea, and the United States of America.

\section*{Data Availability}
All observational data used in this work is publicly available on the MWDD website \citep{dufour2017}, SDSS archive, Pan--STARRS archive, and Gaia archive.

\bibliographystyle{mnras}
\bibliography{references}

\bsp
\label{lastpage}

\end{document}